\documentclass[runningheads]{llncs}
\usepackage{graphicx}
\usepackage[symbol*]{footmisc}
\usepackage{subfigure}
\usepackage{amsmath,amssymb}
\usepackage{booktabs}
\usepackage{textcomp}

\begin{document}

\title{Adaptive Weighting Depth-variant Deconvolution of Fluorescence Microscopy Images with Convolutional Neural Network} 
\titlerunning{  } 


\author{Da He\thanks{The authors contributed equally to this work.} \and
De Cai$^*$ \and
Jiasheng Zhou \and
Jiajia Luo\thanks{Corresponding authors.} \and
Sung-Liang Chen$^\dag$}

\authorrunning{He \emph{et al.}}

\institute{University of Michigan-Shanghai Jiao Tong University Joint Institute, Shanghai Jiao Tong University, China\\
\email{\{da.he, derrick, jiasheng\_zhou, jiajia.luo, sungliang.chen\}@sjtu.edu.cn}}

\maketitle

\begin{abstract}
Fluorescence microscopy plays an important role in biomedical research. The depth-variant point spread function (PSF) of a fluorescence microscope produces low-quality images especially in the out-of-focus regions of thick specimens. Traditional deconvolution to restore the out-of-focus images is usually insufficient since a depth-invariant PSF is assumed. This article aims at handling fluorescence microscopy images by learning-based depth-variant PSF and reducing artifacts. We propose adaptive weighting depth-variant deconvolution (AWDVD) with defocus level prediction convolutional neural network (DelpNet) to restore the out-of-focus images. Depth-variant PSFs of image patches can be obtained by DelpNet and applied in the afterward deconvolution. AWDVD is adopted for a whole image which is patch-wise deconvolved and appropriately cropped before deconvolution. DelpNet achieves the accuracy of 98.2\%, which outperforms the best-ever one using the same microscopy dataset. Image patches of 11 defocus levels after deconvolution are validated with maximum improvement in the peak signal-to-noise ratio and structural similarity index of 6.6 dB and 11\%, respectively. The adaptive weighting of the patch-wise deconvolved image can eliminate patch boundary artifacts and improve deconvolved image quality. The proposed method can accurately estimate depth-variant PSF and effectively recover out-of-focus microscopy images. To our acknowledge, this is the first study of handling out-of-focus microscopy images using learning-based depth-variant PSF. Facing one of the most common blurs in fluorescence microscopy, the novel method provides a practical technology to improve the image quality.
	
\keywords{Adaptive weighting, convolutional neural network, deconvolution, fluorescence microscopy}
\end{abstract}

\section{Introduction}
Fluorescence microscopy is widely used in biomedical applications such as visualizing three-dimensional (3D) structures of cells and tissues \cite{liu2015imaging}. There are two kinds of blurs in fluorescence microscopy: one is caused by the depth-variant microscopic point spread function (PSF) and another by Poisson noise. The former is also associated with the limited depth of field in a fluorescence microscope, which inevitably causes low-quality images in out-of-focus regions (i.e., the defocus blur) especially for thick specimens \cite{murray2011methods}. Therefore, the limitation severely impacts the performance of fluorescence microscopy as shown in Fig. 1. Users suffer from wasting much time for tuning the focus of a fluorescence microscope, whereas some out-of-focus regions in the field of view are still difficult to distinguish due to the surface height variation of a sample itself.

In optics, the defocused image can be modeled as the convolution between the latent clear image and a depth-variant PSF. Depth-variant PSF indicates that PSFs are different for objects at various depths. Thus, for a realistic microscopy image, it is common that some regions are clear whereas others may be blurred if these objects come from different depths of a thick specimen.

\begin{figure}[!tpb]
	\centerline{\includegraphics[width=0.6\columnwidth]{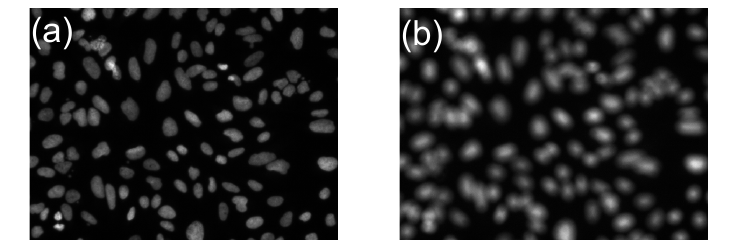}}
	\caption{Two examples of fluorescence microscopy images. (a) An in-focus image. (b) An out-of-focus image.}
	\label{fig1}
\end{figure}



Clear images will be obtained if the above convolution can be reversed. Therefore, deconvolution has been intensively studied to restore the out-of-focus images computationally through non-blind, blind, and semi-blind ways, depending on how much the system's PSF is known. Non-blind deconvolution requires the known PSF from either theoretical estimation or experimental measurement \cite{shaevitz2007enhanced}. However, theoretical estimation of PSF often simply assumes a depth-invariant imaging model as it is really hard to practically measure the depth of an object (i.e., the degree of defocus). This rough approximation naturally leads to a limited improvement for image quality. Besides, some experimental PSF measurements using sensors like wave-front detectors \cite{arines2011partially} are inconvenient and time consuming, which are impractical for biomedical applications. On the other hand, blind deconvolution sidesteps the acquirement of PSF by using only the blurred image \cite{kim2015blind}, yet the deconvolution effect is usually limited due to the lack of system's prior knowledge of optical imaging. Some other effective solutions can be named as semi-blind methods. Instead of directly recovering images, semi-blind methods estimate some variable parameters of a given PSF model, followed by the deconvolution using this semi-predicted model. The imaging information is learned from  the connection between PSF parameters and blurred images. In works about optical and ultrasound imaging \cite{shajkofci2018semi,morin2013semi}, semi-blind deconvolution can increase the overall robustness and generality of image restoration.

Recently convolutional neural network (CNN) has shown its great potentials for image processing, which implements various computer vision tasks such as image classification by ResNet \cite{he2016deep}, object detection by YOLO \cite{redmon2016you}, and instance segmentation by Mask R-CNN \cite{he2017mask}. CNN is a learning-based method and has the power to work for a variety of input images including medical and biological images. CNN-based methods have also been applied to deblur images and remove artifacts. Some large convolution kernels as well as traditional deconvolution schemes are combined in \cite{xu2014deep}. The mapping from blurred images to clear images is learned by CNN to handle adaptive optics retinal images in \cite{fei2017deblurring}.

According to the semi-blind pipeline, the microscopy images can be efficiently deblurred by regressing PSF parameters through CNN and running deconvolution \cite{shajkofci2018semi}. However, in \cite{shajkofci2018semi}, the PSF model assumes the spatially-variant blurring for thin objects, which is inappropriate for thick specimens with depth-variant blurring. Besides, the median filtered patch-wise deconvolution in \cite{shajkofci2018semi} cannot appropriately handle the subtle blurring characteristic in realistic biomedical microscopy, which possibly requires some weighted average operations for blur variations near patch boundaries.  Adaptive deconvolution is an effective way  to restore the original image according to the image local characteristics \cite{yan2012blind,yan2012semi,tsai2014improved}.

In this paper, we therefore proposed adaptive weighting depth-variant deconvolution (AWDVD) with defocus level prediction convolutional neural network (DelpNet) to restore the out-of-focus images in fluorescence microscopy. The main contributions of this article can be listed as follows: we (a) quantitatively estimated the depth parameters using DelpNet , which was initially inspired by \cite{yang2018assessing} but finally reduced the top-$1$ error from 5\% to 1.8\%, (b) firstly deblurred microscopy images with learning-based depth-variant PSFs, which are more common in biomedical practices compared with the previous depth-invariant or spatially-variant works, and (c) compared our special settings with various well-known CNN architectures to give some experiment-based suggestions for processing microscopy images.

The other sections of this article are organized as follows. We introduce our methods in detail in Section \ref{sec:methods}, including our data generation, DelpNet, and AWDVD. Then, the experimental results are presented in Section \ref{sec:results}. Besides the output of the proposed pipeline, we also show special settings of DelpNet and compare it with several main-stream CNN architectures. Results from both AWDVD and depth-invariant deconvolution are also compared in this section. In Section \ref{sec:discussion}, we discuss our contributions and limits. Finally, we conclude in Section \ref{sec:conclusion}.

\begin{figure*}[!tpb]
	\centerline{\includegraphics[width=\columnwidth]{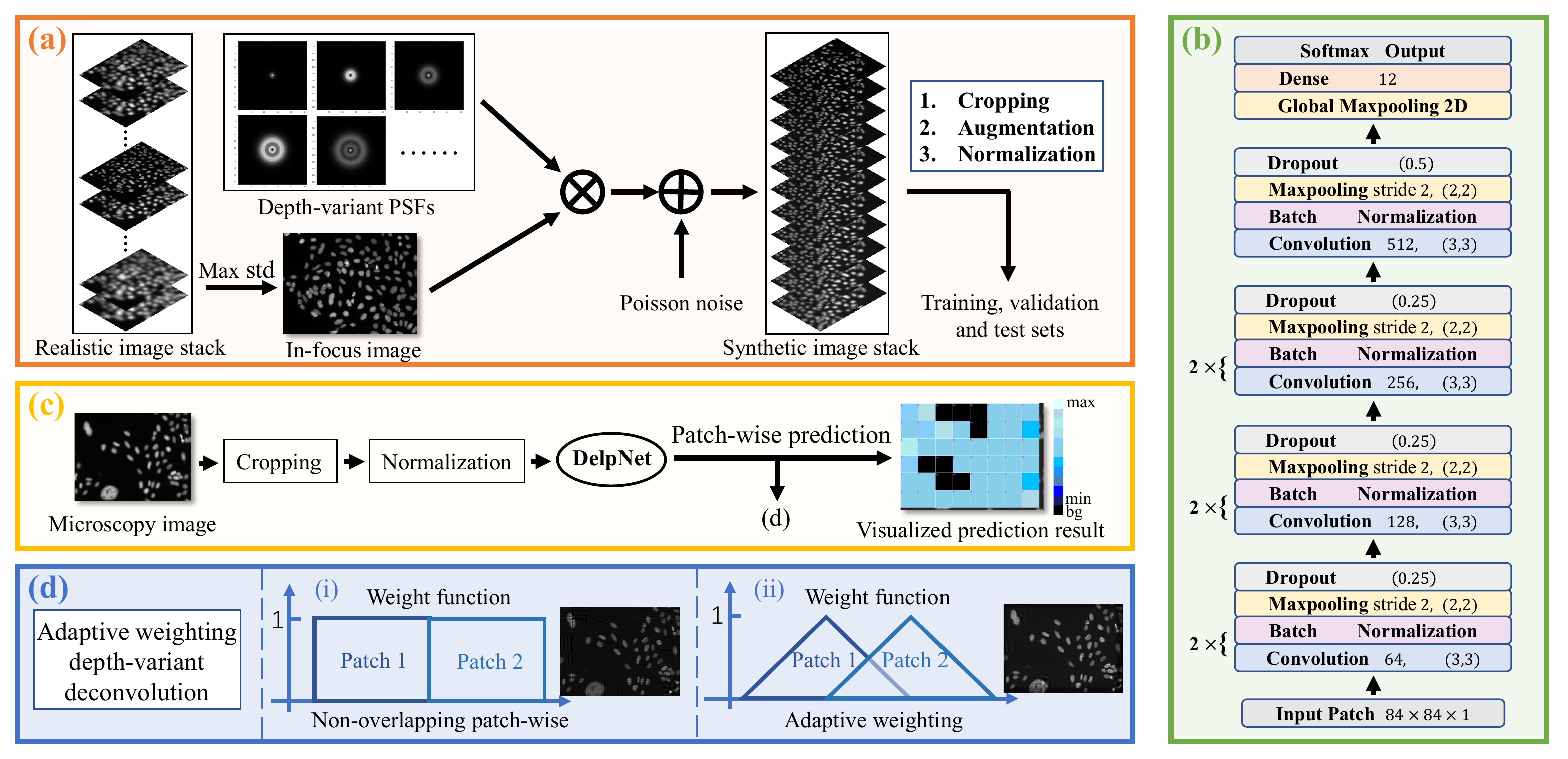}}
	\caption{The overview of the method. (a) The process of generating synthetic datasets. (b) The architecture of DelpNet. (c) An example of the patch-wise defocus level prediction process. The prediction results are expressed in different colors, and the ``bg'' here indicates a background patch. (d) The AWDVD for two cases: (i) the stride is the same as the patch size and it is just the non-overlapping patch-wise deconvolution with rectangular weight function and (ii) the stride is smaller than the patch size and triangular function is used for adaptive weighting.}
	\label{fig2}
\end{figure*}

\section{Methods}
\label{sec:methods}
Firstly, we classified the defocus levels of patches in fluorescence microscopy images by DelpNet. The depth-variant PSF can then be recognized and used in deconvolution for image restoration. For whole image processing, the patch boundary artifacts were eliminated and the deconvolved image quality was also improved in AWDVD. The overview of our pipeline is shown in Fig. 2.

\subsection{Fluorescence Microscopy Images}
In reality, fluorescence microscopy images for cells are similar to Fig. 1(a). However, if most cells are out of focus, their images will be blurred, as shown in Fig. 1(b). This imaging process can be modeled as follows. Assuming lateral shift-invariant, the fluorescence microscopy image $i(x,y;z)$ can be represented as:
\begin{equation}i(x,y;z) = \wp(o \otimes h)(x,y;z),\label{eq1}\end{equation}
where $(x, y; z)$ represents a \emph{2D} function over $x,y$ with a variable $z$ coincident with the optical axis, $\wp$ is the Poisson noise model, $\otimes$ stands for the convolution operator, $o(x,y;z)$ is the imaged object and $h(x,y;z)$ is the depth-variant PSF.

Therefore, we can recover $o(x,y;z)$ if the depth-variant PSF $h(x,y;z)$ can be estimated and used for deconvolution. For an optical or fluorescence microscope, the information of the depth $z$ of the PSF $h(x,y;z)$ varies depending on the depth where the object is placed. That is, the depth $z$ is a parameter that cannot be determined from the specification of the microscope. Fortunately, the depth $z$ can be inferred from the captured image $i(x,y;z)$, as indicated in Eq. (1). This is why we train the following DelpNet to estimate depth information of an image. To prepare training data as supervised learning requires, we generated 11 defocus levels of synthetic images from realistic experimental focal images. We avoided directly using realistic defocused images to form the training set, because it is usually hard to perfectly model a microscopy system's PSF and accurately control defocus levels as labels in experiments. Instead, using an approximate PSF model with accurately predicted defocus levels by virtue of synthetic images as the data to train DelpNet still contributed a lot to the subsequent deconvolution even for realistic images for more practical applications. Thus, we generated synthetic training data based on realistic images to train DelpNet, as shown in Fig. 2(a), and applied the trained model to realistic images with unknown defocus levels , as shown in Fig. 2(c).

We used fluorescence images of U2OS cells with Hoechst stain in BBBC006 dataset \cite{ljosa2012annotated} for experimentation. The dataset acquired by ImageXpress Micro automated cellular imaging system (Molecular Devices, CA) consists of image stacks for every specimen at different depths. The 381 in-focus images of size $696 \times 520$ were first selected based on the maximum standard deviation of image intensity from 381 image stacks. They are regarded as the reference in-focus images. Synthetic defocused image dataset was then obtained as shown in Fig. 2(a) by convolving the selected in-focus images with the following depth-variant PSFs of 11 defocus levels \cite{yang2018assessing,Born1999Elements}:
\begin{equation}  h(x,y;z) =  
\left| \int_0^1 \! J_0 \! \left( \! k \!\frac{NA}{n} \! \sqrt{ \! x^2 \! + \! y^2} \! \rho \! \right) exp \! \left( \! - \! \frac{1}{2}jk\rho^2 z \! \frac{NA^2}{n^2} \! \right) \! \rho d\rho\right|^2 \! , \label{eq2} \end{equation}
where $J_0$ is the Bessel function of the first kind, order zero, $k = 2\pi / \lambda$ is the wavenumber with the wavelength $\lambda$ as $451\ nm$, $NA$ is $0.5$, and $n = 1.0$ is the refractive index. The defocus level has the increment of 2 \textmu m along $z$ optical axis and level 0 corresponds to the in-focus images (i.e., no synthetic degradation). After randomly splitting the synthetic dataset into training, validation and test sets with the ratio of $0.75 : 0.15 : 0.1$, the images of size $696 \times 520$ in three sets were then randomly cropped to image patches of size $84 \times 84$. Each whole image was cropped for 20 times. We assigned some patches the label ``bg'', whose maximum pixel values are smaller than 230 and the maximum difference among pixel values are smaller than 30 considering 16-bit images, indicating that the patch is almost full of background noise. This special augmentation is helpful because it can partially avoid providing DelpNet with meaningless interference. After the ``bg'' augmentation, we had 12 candidate types of image patches. Linear normalization was applied to all image patches to scale their value ranges to 0--1. The patches with defocus level labels were then used to train DelpNet for classification as well as the following analysis.

\subsection{DelpNet}
As shown in Fig. 2(b), DelpNet includes 7 convolutional layers. The shape of the input layer was set to match the input patch size. Dropout \cite{hinton2012improving} and BatchNormalization \cite{ioffe2015batch} layers were added to avoid overfitting. Every BatchNormalization layer was followed by a ReLU activation function.  GlobalMaxPooling layer \cite{lin2013network} was utilized to transform the convolutional feature maps to a feature vector before the output layer. We have 12 classification outputs for 11 defocus levels together with the extra ``bg'' label.

There are some special settings in DelpNet. Firstly, we chose a relatively plain CNN architecture instead of residual learning style like ResNet \cite{he2016deep} or multi-scale feature fusion design like Inception network \cite{szegedy2016rethinking}, as plain architectures surprisingly performed better in our experiments. Secondly, it is uncommon to densely put Batch Normalization layers and Dropout layers together, but we found this strategy useful for improving the performance and reducing overfitting. In addition, we modified the default momentum value of Batch Normalization \cite{ioffe2015batch} layers in Tensroflow from 0.99 to 0.60, which is also unusual but helpful for convergence for our dataset. All of these strategies will be further compared in Section \uppercase\expandafter{\romannumeral3}.

As for implementation, DelpNet was built using Keras framework with Tensorflow \cite{abadi2016tensorflow} backend. We set Adam optimizer \cite{kingma2014adam} learning rate of $6e-5$ and decay of $5e-6$.  With batch size of 128, it was then trained and evaluated on a single Nvidia Titan Xp GPU. Cross-entropy loss was applied to this multiclassification problem.

\subsection{AWDVD}
Assuming Poisson noise for the imaging model, Richardson-Lucy deconvolution \cite{lucy1974iterative,richardson1972bayesian} is a widely-used iterative algorithm to restore the original object image with known PSF:
\begin{equation}
O_{t+1}'(x,y;z) = 
\left[ \frac{i(x,y;z)}{h(x,y;z) \otimes o_t'(x,y;z)} \otimes h(-x,-y;-z) \right] o_t'(x,y;z),
\label{eq3}\end{equation}
where $t$ represents the iterative index, $o'(x,y;z)$ is the estimated object profile and the initial guess of $o_0'(x,y;z) $ is set as $i(x, y; z)$. The depth-variant PSF based on the predicted defocus level by DelpNet was assigned for deconvolution for each image patch of size $84 \times 84$.

For a whole large image, patch-wise prediction of defocus levels was firstly applied to image patches ($84 \times 84 $) with certain stride, and then, AWDVD followed, as elaborated below. For case (i) in Fig. 2(d), because strides were simply set as the patch size and a rectangular weight function was naturally applied, the non-overlapping patch-wise deconvolution way cannot preserve the continuity at the patch boundary (i.e., patch boundary artifacts). Besides, the captured microscopy image within an image patch usually presents both in-focus and out-of-focus cells. Thus, it is preferred to know a more local defocus information within an image patch, which cannot be realized by case (i). For AWDVD of case (ii), a triangular function with base width the same as the patch size (i.e. 84 pixels) was used as the weight function to eliminate the patch boundary artifacts, and more local defocus levels can thus be used for deconvolution. The height of triangular function was correspondingly scaled by the ratio of the stride over half of the patch size. When half of the patch size (i.e., 42) is divisible by the strides (1, 2, 3, 6, 7, 14, 21, and 42), the weight function shall have the top hat shape just like a rectangular function. If the stride is just half of the patch size, then it is just the bilinear interpolation of adjacent patches \cite{thiebaut2016spatially}. In addition, patches with predicted ``bg'' category were directly fed to weight function without deconvolution.

\section{EXPERIMENTAL RESULTS AND ANALYSES}
\label{sec:results}
We used the held-out test set to evaluate the defocus level prediction performance and compare strategies in the following subsections A, B, and C . The overall results sequentially combining DelpNet and AWDVD were analyzed in subsection D.

\subsection{Results of DelpNet}
After 1000 epochs of training, the evaluation results of DelpNet on test set are shown in Fig. 3, which shows the distribution of prediction results using a confusion matrix.

Generally, our DelpNet method can identify the defocus levels with an accuracy of 98.2\%, and the precision, recall as well as f1-score are all above 98\%. Besides, from the confusion matrix in Fig. 3, we can easily find that the wrong predictions mostly locate in the neighbors of the ground truth, which means that most wrong predictions differ from the ground truth slightly, and thus, the subsequent deconvolution with these wrong predictions can still contribute to image restoration to some extent.

\begin{figure}[!htpb]
	\centerline{\includegraphics[width=0.6\columnwidth]{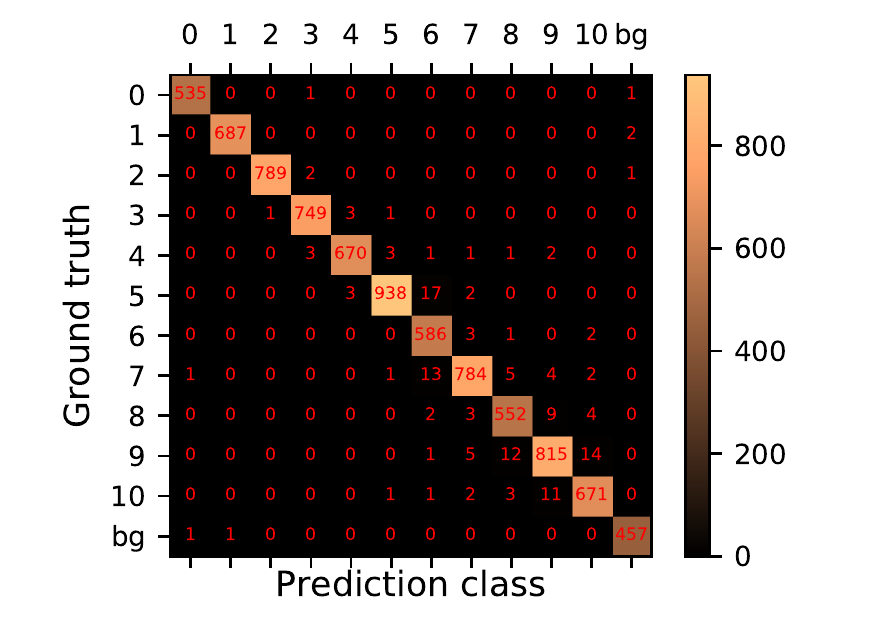}}
	\caption{Evaluation result of DelpNet. The confusion matrix shows the distribution of prediction results of DelpNet on the held-out test set.}
	\label{fig3}
\end{figure}

\subsection{Analysis of Special Settings in DelpNet}
We manually set an extra label ``bg'' to split patches with almost full background from normal defocused patches. To analyze the value of this strategy, we trained the almost same DelpNet with two settings: with ``bg'' label and without ``bg'' label. The only difference existed in the final dense layer, which has 12 neurons and 11 neurons for cases with ``bg'' label and without ``bg'' label, respectively. The evaluation results are listed in Table 1.

\begin{table}[h]
	\centering
	\caption{Evaluation Results of the Existence of ``bg'' Label}
	\label{table1}
	\begin{tabular}[c]{ccccc}
		\toprule
		Setting & Accuracy & Precision & Recall & F1-score \\
		\midrule
		\emph{Without ``bg''} & 96.8\% & 96.6\% & 96.9\% & 96.7\% \\
		\emph{With ``bg''} & \textbf{98.2\%} & \textbf{98.2\%} & \textbf{98.3\%} & \textbf{98.3\%} \\
		\bottomrule
	\end{tabular}
\end{table}

As shown in Table 1, the ``bg'' label strategy worked well with accuracy improvement from 96.8\% to 98.2\%, which is equivalent to the decrease of top-$1$ error from 3.2\% to 1.8\%. As for reasons, we think using ``bg'' label is very important to prevent CNN from excessively fitting the background. Besides, synthetic operations have also been done for ``bg'' patches, whereas the background areas in realistic defocused images almost contain random noise only, instead of the ``convolved noise''. Thus, this strategy avoids learning defocus levels from these fictional features (i.e., convolved noise) and contributes to the applicability for realistic out-of-focus images.

\begin{figure}[!hbtp]
	\centerline{\includegraphics[width=0.6\columnwidth]{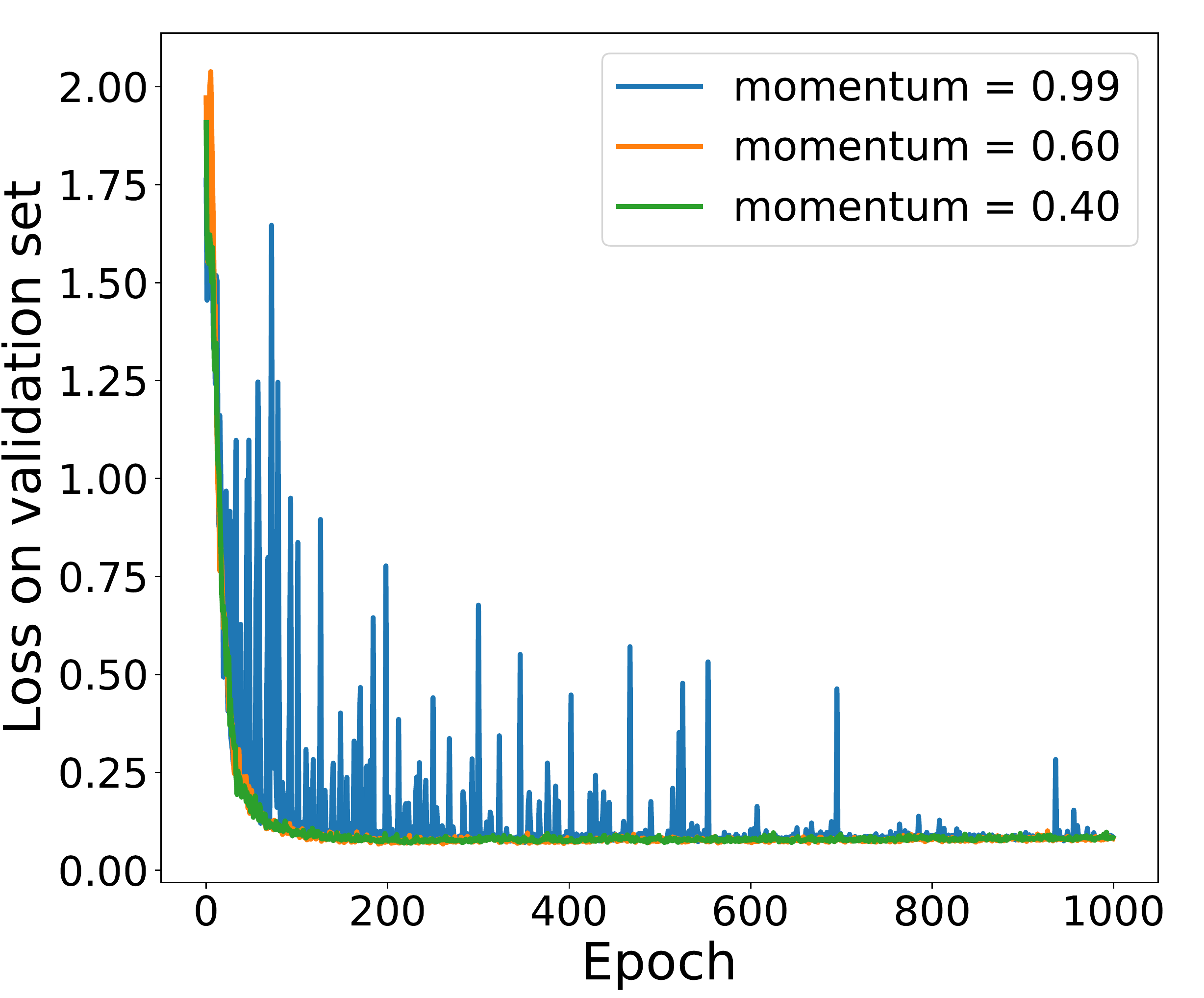}}
	\caption{The validation loss curves of three Batch Normalization momentum parameters. The blue curve showing momentum of 0.99 is disordered.}
	\label{fig4}
\end{figure}
Different from most works that apply Batch Normalization with the momentum parameter of a relatively large value (e.g.,  default 0.99 in \cite{abadi2016tensorflow}), we assigned this parameter 0.60. This is because we found large Batch Normalization momentum parameters easily lead to oscillations on validation set while training. As shown in Fig. 4, a severe oscillation occurred on the curve of validation set. This phenomenon probably came from the characteristics of fluorescence microscopy images. On the other hand, relatively small values of the momentum would not ruin the performance in our experiments. Thus, we chose a relatively small value of 0.60 to help the training converge steadily.

Besides the two strategies above, we also densely used Batch Normalization layers with Dropout layers. This combination is powerful as well. For fluorescence microscopy images, there are much more background noises than meaningful cell signals. As a result, more strategies of easing overfitting are needed, compared with processing natural photos. Fig. 5 shows the cumulative match characteristic curves of DelpNet and DelpNet without the three strategies, respectively. The top-$k$ accuracies in Fig. 5 indicate that each of the proposed settings is instrumental in improving the performance.

\begin{figure}[!htpb]
	\centerline{\includegraphics[width=0.6\columnwidth]{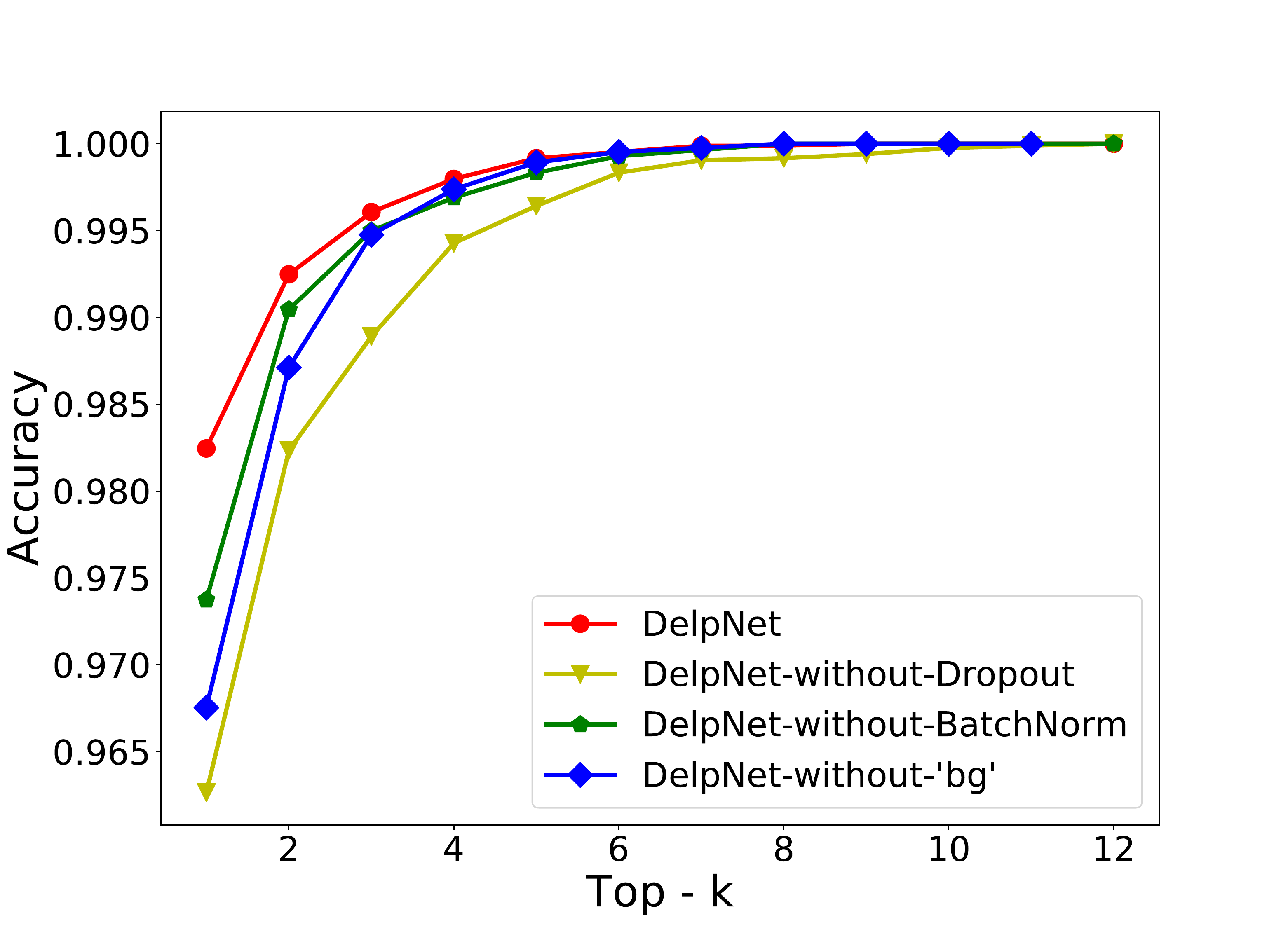}}
	\caption{Cumulative match curves of different settings. Each node represents the corresponding top-$k$ accuracy of a specified training strategy (e.g., the most left blue node indicates that if we train DelpNet model without ``bg'' label, the top-$1$ accuracy on the test set is about 96.8\%.). The variable $k$ is in the range of 1--11 for the blue curve and the range of 1--12 for other curves.}
	\label{fig5}
\end{figure}

\subsection{Comparison among Architectures}

Besides the above validation of our special strategies in the proposed DelpNet, we also compared the performance of DelpNet with various well-known representative CNN architectures in Fig. 6 and Table 2. Specifically, VGG16 \cite{simonyan2014very} represents the plain CNN architectures without any concatenations or additions between feature maps. Inception\_v3 \cite{szegedy2016rethinking} simultaneously applies different convolutional kernels and concatenates the feature maps together for multi-scale fusion. ResNet18 \cite{he2016deep} is a typical residual learning network with relatively fair layer numbers compared with its deep versions. Finally, MobelNet\_v2 \cite{sandler2018mobilenetv2} is a light work proposed to implement CNN applications on mobile devices. Compared with their original architectures, we only adjusted the input and output formats to fit our dataset.

\begin{figure}[!t]
	\centerline{\includegraphics[width=0.6\columnwidth]{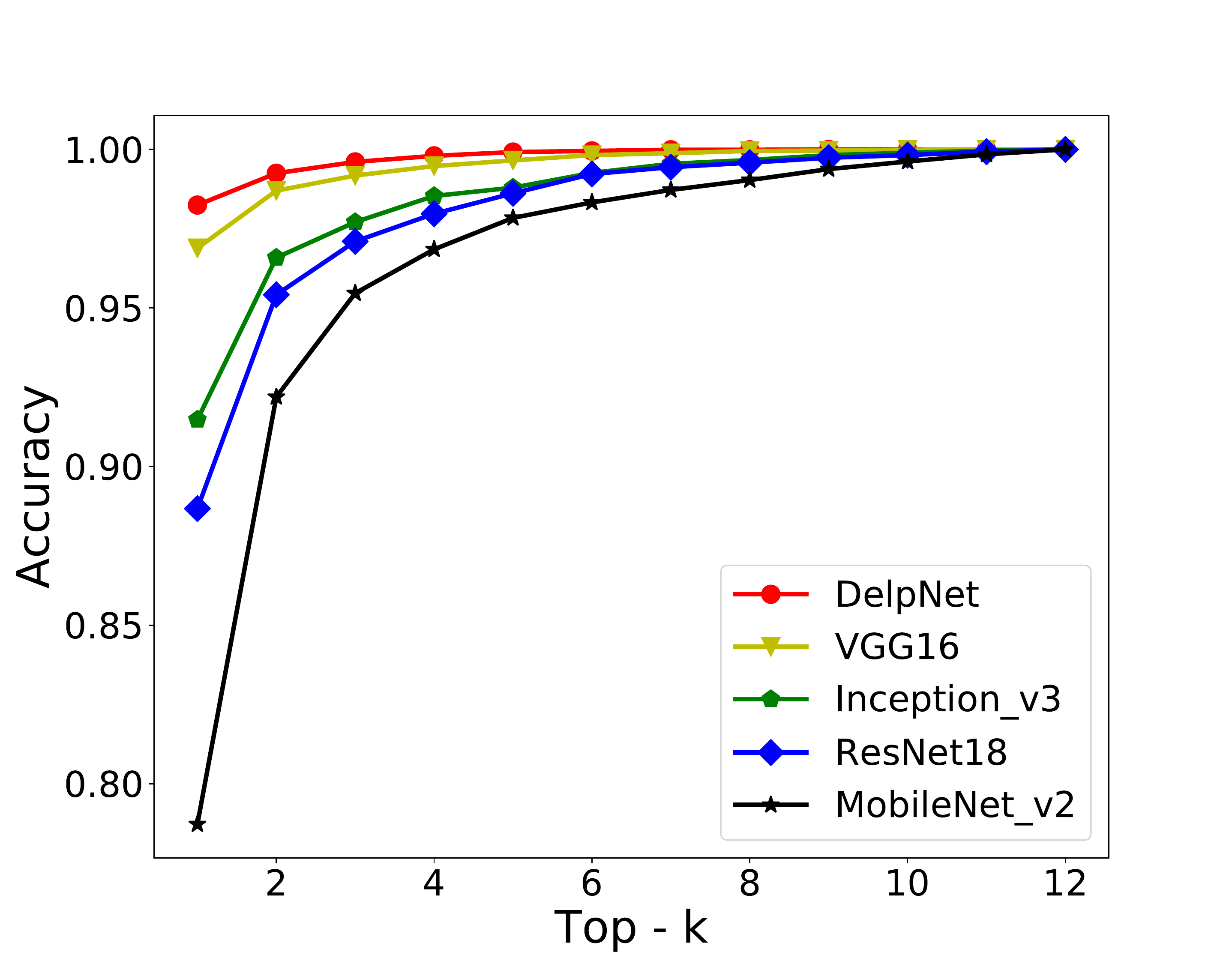}}
	\caption{Cumulative match curves of trained DelpNet and other 4 representative CNNs. Each node represents the corresponding top-$k$ accuracy of a CNN architecture (e.g., the most left blue node indicates that using the trained ResNet18 model, the top-$1$ accuracy on the test set is about 88.7\%.). All the architectures were retrained by us with slight modifications for input and output formats.}
	\label{fig6}
\end{figure}

\begin{table}[!tbp]
	\centering
	\caption{Evaluation Results of Various CNN Architectures}
	\label{table2}
	\begin{tabular}[c]{cccccc}
		\toprule
		Net & \# params & Accuracy & Precision & Recall & F1-score \\
		\midrule
		\emph{VGG16} & 39.9M & 96.9\% & 96.9\% & 97.0\% & 96.9\% \\
		\emph{Inception\_v3} & 21.8M & 91.5\% & 91.5\% & 91.7\% & 91.6\% \\
		\emph{ResNet18} & 12.6M & 88.7\% & 88.7\% & 89.1\% & 88.8\% \\
		\emph{MobileNet\_v2} & \textbf{2.3M} & 78.7\% & 79.0\% & 79.6\% & 79.1\% \\
		\emph{DelpNet} & \textbf{2.3M} & \textbf{98.2\%} & \textbf{98.2\%} & \textbf{98.3\%} & \textbf{98.3\%} \\
		\bottomrule
	\end{tabular}
\end{table}

\begin{figure*}[!htb]
	\centerline{\includegraphics[width=\columnwidth]{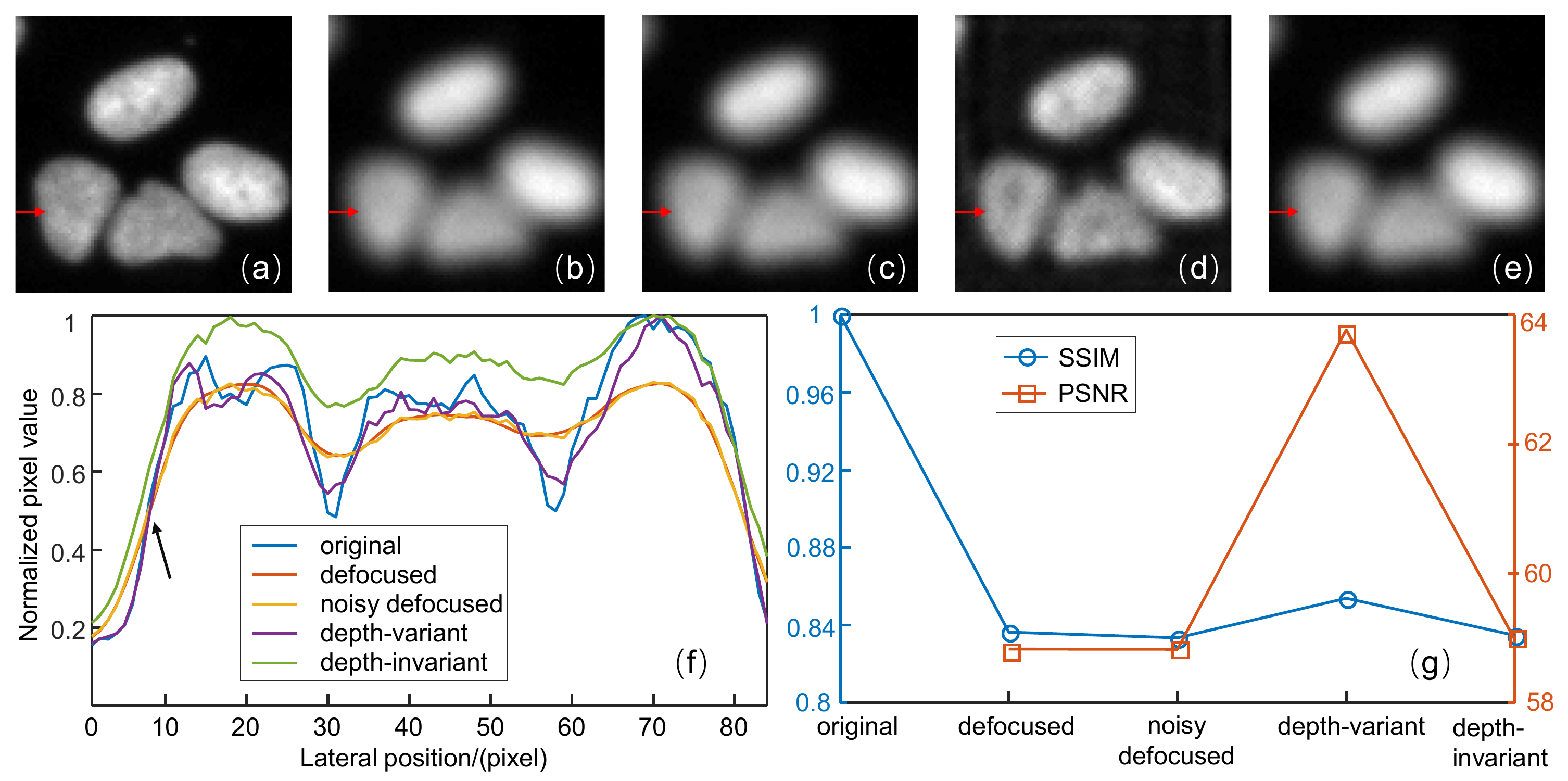}}
	\caption{An image patch for (a) original in-focus image, (b) defocused image, (c) noisy defocused image, (d) depth-variant deconvolved image, and (e) depth-invariant deconvolved image. (f) The 1D profile of the above five images along the red arrow direction. (g) The SSIM and PSNR values of the above five images, taking (a) as the comparison reference.}
	\label{fig7}
\end{figure*}

Obviously, according to Fig. 6 and Table 2, DelpNet outperforms all the other evaluated CNN architectures. Thus, it is necessary for us to design such a special CNN instead of directly adopting an existing well-known architecture. Besides, according to Table 2, there are only 2.3M parameters in DelpNet, which is helpful for integrating DelpNet into embedded systems like an intelligent microscope.


\subsection{Results and Comparisons of AWDVD}

The trained DelpNet was then used for defocus level prediction and the subsequent deconvolution. An image patch with 3 adjacent U2OS cells is shown in Fig. 7. The PSF with a random defocus level blurred the in-focus image, resulting in Fig. 7(b). Poisson noise was added to the defocused image as shown in Fig. 7(c) of the noisy defocused image. Images by depth-variant deconvolution using the PSF of the predicted defocus level 4 and conventional depth-invariant deconvolution with the in-focus PSF (i.e., defocus level 0) are shown in Figs. 7(d) and 7(e), respectively. In Fig. 7(f), we compared the 1D profiles along the red arrows (the whole 84 pixels) in Figs. 7(a)--7(e). As shown in Fig. 7(f), the PSF and Poisson noise blurred the structure of the original image. The depth-invariant deconvolution can barely restore the original fine structure, whereas depth-variant deconvolution recovered the profile quite well with contrast improvement. The edge after the depth-variant deconvolution was also enhanced, as indicated by the black arrow in Fig. 7(f). For quantitative comparison, the peak signal-to-noise ratio (PSNR) and structural similarity index (SSIM) \cite{wang2004image} were plotted in Fig. 7(g). The PSNR in depth-variant deconvolved image was improved to 63.8 dB from about 59.0 dB in the defocused, noisy defocused and depth-invariant deconvolved images. The SSIM was improved to 0.86 from 0.83 in the noisy defocused image. 

\begin{figure}[!htbp]
	\centerline{\includegraphics[width=\columnwidth]{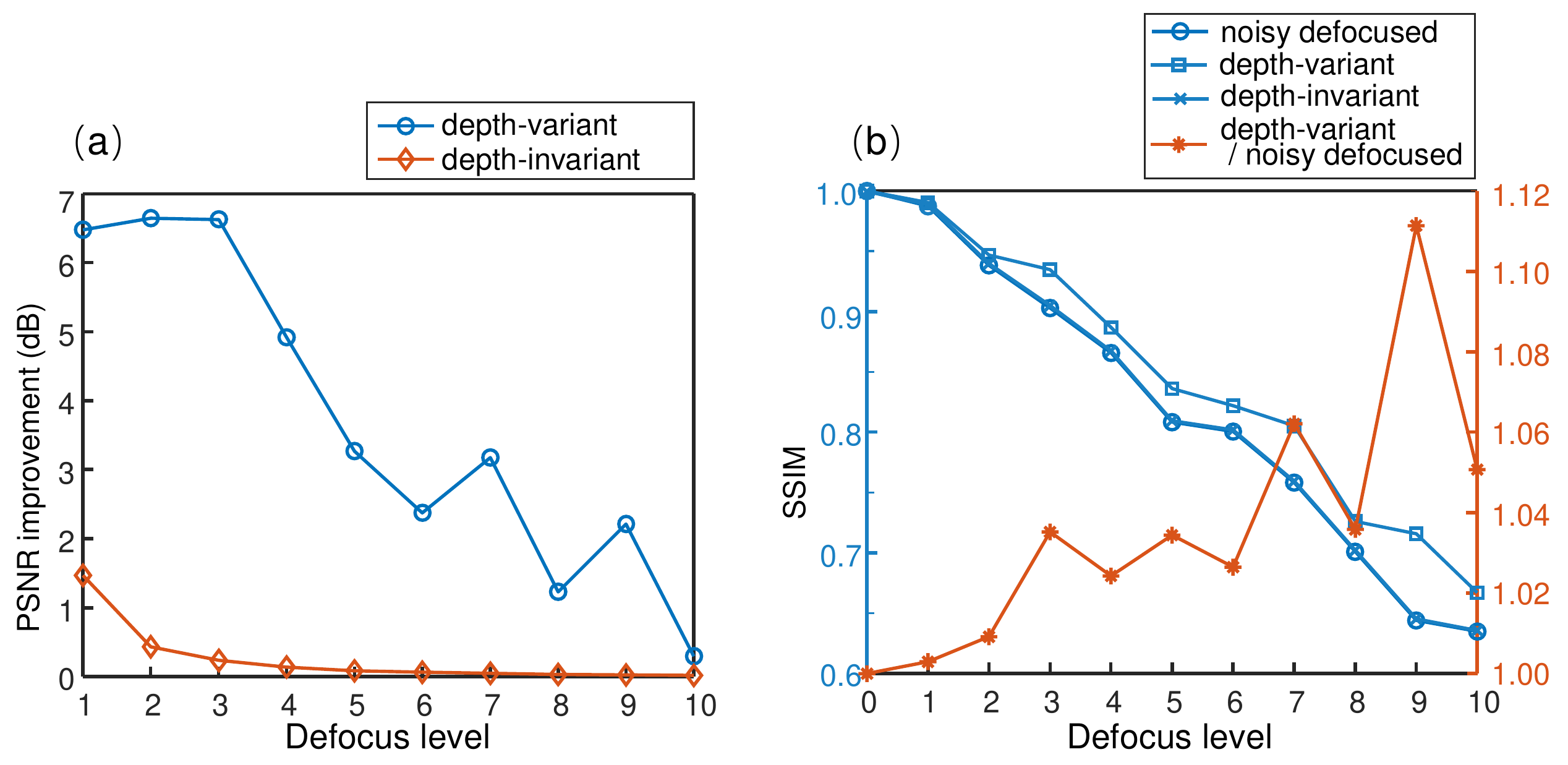}}
	\caption{(a) PSNR improvement for depth-variant and depth-invariant deconvolved images. (b) SSIM of the noisy defocused, depth-variant deconvolved, and depth-invariant deconvolved images; the SSIM ratio (the orange curve and text), defined as the SSIM of the depth-variant deconvolved image over that of the noisy defocused one.}
	\label{fig8}
\end{figure}

The PSNR and SSIM improvements were calculated for each of the 11 defocus levels using about 45 image patches for each defocus level for statistical purposes. The results are shown in Figs. 8 (a) and 8(b) for the PSNR and SSIM, respectively. The PSNR improvement was compared with the noisy defocused images. The depth-variant deconvolution and depth-invariant deconvolution attained PSNR improvement of 0.3--6.6 dB and 0--1.5 dB, respectively. That is, the depth-variant deconvolution performed much better in PSNR improvement. As the defocus level grows, the PSNR improvements show a decreasing trend for both depth-variant and depth-invariant deconvolutions.  The SSIM was calculated by choosing the original in-focus patches as reference. The SSIM decreased for the noisy defocused and deconvolved images (depth-variant or depth-invariant) as defocus level increased.  Because larger defocus level with intensity-dependent Poisson noise resulted in lower SNR, leading to more difficulties for recovery. The SSIM of the depth-variant deconvolution ranged from 0.66 to 1.0, which  is higher than that of the noisy defocused and depth-invariant deconvolution cases. Furthermore, SSIM improvement was evaluated by checking the SSIM ratio, which is defined as the SSIM of the depth-variant deconvolved image over that of the noisy defocused one. Overall, the SSIM ratio increased with the defocus level, achieving improvement of 0--11\%.

Fig. 9 shows a representative realistic whole image by AWDVD of two cases illustrated in Fig. 2(d). Figures 9(a), 9(b), and 9(c) shows realistic image, case (i) in Fig. 2(d), and AWDVD of case (ii), respectively. As can be seen in Fig. 9(b), non-overlapping patch-wise deconvolution suffered from patch boundary artifacts. For AWDVD of case (ii), the stride was set as half of the patch size since the cells were not densely distributed. That is, the weighting processing was bilinear interpolation. As shown in Fig. 9(c), patch boundary artifacts disappeared, and increased smoothness of the whole image can be clearly observed. For better visualization, the image blocks indicated by red dashed boxes in Figs. 9(a)--9(c) were zoomed-in in Figs. 9(d)--9(f), respectively, for close comparison.

\begin{figure}[!htbp]
	\centerline{\includegraphics[width=0.6\columnwidth]{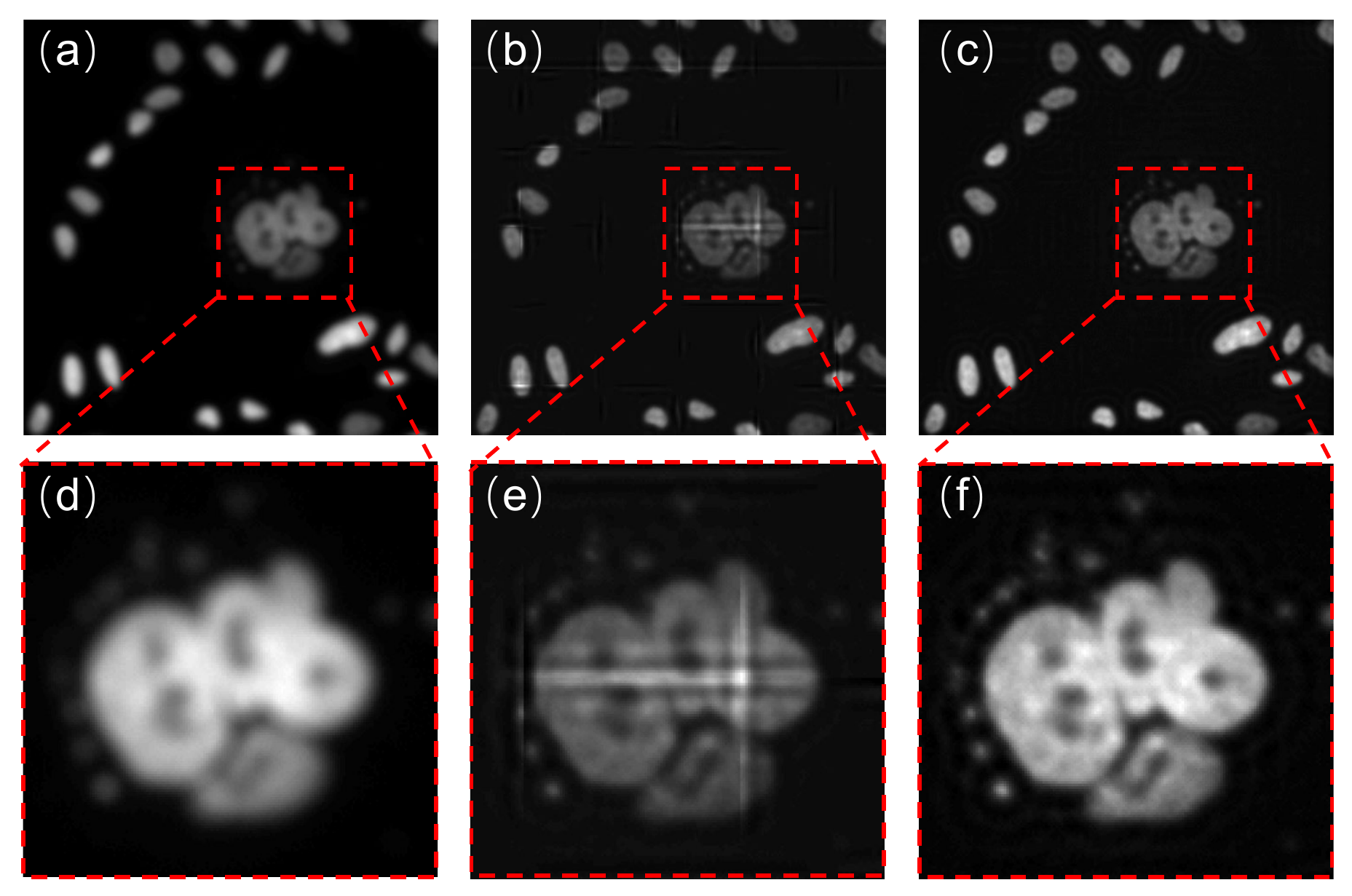}}
	\caption{Whole image results for (a) realistic image, (b) AWDVD of case (i) in non-overlapping patch-wise deconvolution, and (c) AWDVD of case (ii) in a bilinear interpolation way. (d)--(f) are the zoomed-in image blocks in the red dashed boxes in (a)--(c) respectively.}
	\label{fig9}
\end{figure}

\section{Discussion}
\label{sec:discussion}
We proposed AWDVD with DelpNet to handle out-of-focus fluorescence microscopy images. Various experiments and comparisons have been done to prove the effectiveness of this approach.

Compared with \cite{yang2018assessing}, although part of our work (i.e., DelpNet) and \cite{yang2018assessing} both aimed to estimate the defocus levels, our DelpNet used specially designed CNN architecture which achieved better performance.  With the same patch size and the number of defocus levels as those in \cite{yang2018assessing}, our model achieved tremendous accuracy improvement from 95\% in \cite{yang2018assessing} to 98.2\%. The improvement could be attributed to the fact that we appropriately design a plain architecture and introduced ``bg'' label to avoid overfitting. In addition, we further applied the defocus level prediction algorithms to recover defocused images, which is a significant application.

Compared with \cite{shajkofci2018semi}, we focused on the depth-variant blurring instead of the spatially-variant one. Thus, we built a depth-variant PSF model and applied DelpNet to estimate the depth information, which is more concise, compared with those estimated parameters in \cite{shajkofci2018semi}. Besides, we proposed AWDVD method to recover images from patches instead of a simple median filtering in \cite{shajkofci2018semi}. As a result, we were able to apply smaller strides for patch cropping and eliminated the patch boundary artifacts results from the variation of defocus levels. Even more, AWDVD has the potential to handle pixel-wise deconvolution if patch stride of 1 is adopted.

Besides, the analysis of special settings and comparisons between DelpNet and other architectures show some general inspirations for similar cell images. Firstly, for a specified dataset, plain CNN could have better performance than residual learning or multi-scale fusion CNNs. Besides, for biomedical images with much background like microscopy cell images, it is helpful to try various strategies to avoid overfitting on background noises. Strategies such as splitting ``bg'' label and simultaneously using Batch Normalization \cite{ioffe2015batch} with Dropout \cite{hinton2012improving} can be considered.

There are some limits of our dataset for further improvements. From the original image stack in BBBC006 \cite{ljosa2012annotated}, we took the image with the maximum standard deviation as the focal one in this stack to generate defocused images as well as the following patches. However, these selected images are not strictly in-focus as there are still some slight out-of-focus cells due to the samples' thicknesses. This will bring interferences and surely hinder the performance. Furthermore, if we have strictly focal images as the basis for synthesis, we could accurately label pixel-wise defocus levels and even implement pixel-wise defocus level prediction.

\section{Conclusion}
\label{sec:conclusion}
In this paper, we proposed a semi-blind method to handle out-of-focus fluorescence microscopy images. Firstly, we proposed DelpNet to estimate the defocus level of an image patch. The identification accuracy of 98.2\% was achieved, which is higher than the accuracy of 95\% in \cite{yang2018assessing} based on the same original dataset. We analyzed several special strategies, which could be useful strategies for processing other biomedical images with similar sparse characteristics. Then, we conducted AWDVD for image restoration in fluorescence microscopy with the depth-variant PSF's parameter predicted by DelpNet. Maximum PSNR improvement of 6.6 dB and SSIM improvement of 11\% were obtained for image patches of 11 defocus levels. Besides, AWDVD can handle patch boundary artifacts after decreasing patch cropping strides, and thus improve deconvolved image quality.

\section*{Acknowledgement} 
This work was support by National Natural Science Foundation of China (NSFC) grants 31870942 and 61775134.\\

\bibliographystyle{splncs}
\bibliography{refs}

\end{document}